\documentclass[iop,apj,tighten,twocolappendix,numberedappendix]{emulateapj}
\usepackage{apjfonts}

\usepackage{graphicx}
\usepackage{amsmath}
\usepackage{amssymb}
\usepackage{color}
\usepackage{mathtools}
\usepackage[breaklinks,colorlinks,citecolor=blue,linkcolor=red]{hyperref} 
\usepackage[all]{hypcap}

\newcommand\msun{\, \rm M_\odot}

\def\apj{ApJ}%
\def\apjl{ApJL}%
\def\mnras{MNRAS}%
\begin{document} 

\title{Millisecond pulsars and the gamma-ray excess in Andromeda}
\author{Giacomo Fragione\altaffilmark{1}, Fabio Antonini\altaffilmark{2} and Oleg Y. Gnedin\altaffilmark{3}}
\affil{$^1$Racah Institute for Physics, The Hebrew University, Jerusalem 91904, Israel}
\affil{$^2$STFC E. Rutherford fellow. Faculty of Engineering and Physical Sciences, University of Surrey, Guildford, Surrey, GU2 7XH, UK}
\affil{$^3$Department of Astronomy, University of Michigan, Ann Arbor, MI 48109, USA}


\begin{abstract}
The Fermi Gamma-Ray Space Telescope has provided evidence for diffuse gamma-ray emission in the central parts of the Milky Way and the Andromeda galaxy. This excess has been interpreted either as dark matter annihilation emission or as emission from thousands of millisecond pulsars (MSPs). We have recently shown that old massive globular clusters may move towards the center of the Galaxy by dynamical friction and carry within them enough MSPs to account for the observed gamma-ray excess. In this paper we revisit the MSP scenario for the Andromeda galaxy, by modeling the formation and disruption of its globular cluster system. We find that our model predicts gamma-ray emission $\sim 2-3$ times larger than for the Milky Way, but still nearly an order of magnitude smaller than the observed Fermi excess in the Andromeda. Our MSP model can reproduce the observed excess only by assuming $\sim 8$ times larger number of old clusters than inferred from galaxy scaling relations. To explain the observations we require either that Andromeda deviates significantly from the scaling relations, or that a large part of its high-energy emission comes from additional sources.
\end{abstract}

\keywords{gamma-rays: galaxies --- gamma-rays: diffuse background --- pulsars: general --- galaxies: star clusters: general --- Galaxy: centre --- Galaxy: kinematics and dynamics}

\section{Introduction}
\label{sect:intro}

The gamma-ray luminosity of star-forming galaxies has been under scrutiny for a long time since its study may provide important clues to the acceleration mechanisms of cosmic rays and their transport through the interstellar medium, and constrain the star formation rate as well as the gas and metallicity content of a galaxy. Thanks to the Large Area Telescope instrument on board of the Fermi Gamma-Ray Space Telescope (Fermi-LAT), new high-quality data from $20$~MeV to over $300$~GeV have been available to study the high-energy physics \citep{atw09}. These data have revealed peculiarities of the gamma-ray emission from the inner region of our Galaxy, the so-called Fermi Bubbles -- large structures extending up to 8 kpc away from the Galactic plane \citep{ack14}.

Analyses of the diffuse gamma-ray emission also found a spherically-symmetric excess around the Galactic Centre, peaking at $\sim 2$~GeV and extending out to $\sim 3\,$kpc from the centre \citep{aba14,cal15,lee15,ajel16}. Two main explanations have been proposed for the observed excess, based mainly on similarity with the radial distribution and energy spectrum of the emission. A possibility is that the excess is a product of dark matter annihilation \citep{cal15}. Alternatively, the emission could be due to thousands of unresolved MSPs \citep*{bra15,bar16,arc17,fao18,fpb18}.

Besides the Milky Way, seven external star-forming galaxies have been observed by Fermi in gamma rays, including the Small and Large Magellanic Cloud and the Andromeda galaxy \citep{acker12}. The latter is of particular interest since it is the only other large spiral with a prominent bulge which is close enough that the disk and bulge can be resolved as separate components. Its galactic nucleus harbors a supermassive black hole and a central blue cluster (P3) surrounded by two overdensities of stars (P1 and P2), which reside on either side of P3 with a separation of $\sim 1.8$ pc \citep{bender2005,lauer2012}. \citet{acker17} reported the detection of diffuse gamma-ray emission on the order $\sim 2.8 \times 10^{38}$ erg s$^{-1}$, that extends up to $\sim 5$ kpc from Andromeda's center, with the significance of spatial extent at the $4\sigma$ level. Its morphology is not well constrained and can be described either by a uniform disk or a Gaussian distribution. Compared to the Milky Way's excess, the Andromeda excess is about one order of magnitude larger. Moreover, this emission does not correlate with regions rich in gas, and its spectrum is consistent with a simple power law or with a truncated power-law with an exponential cut-off in the GeV range. The latter closely resembles the MSP spectral templates. As in the Galactic case, there have been claims for both the MSP and dark matter-annihilation origin of the Andromeda's diffuse emission \citep{mcdaniel18}. \citet{acker17} suggested that, if MSPs are responsible for the emission, the $\sim 4-10$ times higher flux in Andromeda could be attributed to the correspondingly higher number of globular clusters in that galaxy \citep{barm01,gall07}. Recently, \citet{eckn17} proposed that the emission comes from an unresolved population of MSPs formed \textit{in situ}.

In this paper, we revisit the MSP scenario in the Andromeda galaxy. We model the formation and disruption of Andromeda globular clusters across all cosmic time, starting from redshift $z=3$ to the present time, and calculate the amount of MSPs deposited in the Andromeda bulge as a consequence of cluster disruption, while accounting also for the spin-down of the MSPs due to magnetic-dipole braking.

This paper is organized as follows. In Section 2, we describe the semi-analytical model we used to generate and evolve the primordial population of globular clusters. In Section 3, we show that our fiducial model underestimates the measured Andromeda excess. Finally, in Section 4, we discuss the implications of our findings and summarize our conclusions.

\section{Globular Cluster evolution}
\label{sect:gcev}

In this section, we discuss the equations used to evolve the globular cluster (GC) population; for details see \citet{gne14}. We assume that the cluster formation rate was a fraction $f_{\mathrm{GC},i}$ of the overall star formation rate
\begin{equation}
\frac{dM}{dt}=f_{\mathrm{GC},i}\frac{dM_{*}}{dt}\ .
\label{eqn:fgci}
\end{equation}
We assume that all clusters formed at redshift $z=3$ and calculate their subsequent evolution for $11.5$ Gyr. The initial mass of the clusters is drawn from a power-law distribution
\begin{equation}
\frac{dN}{dM}\propto M^{-2},\ \ \ \ M_{\min}<M<M_{\max}\ .
\label{eqn:gcmassini}
\end{equation}
We set $M_{\min}=10^4\msun$ and $M_{\max}=10^7\msun$. 

After formation, we evolve the GC masses by taking into account mass loss via stellar winds and the removal of stars by the galactic tidal field. The mass loss is modeled assuming a \citet{kro01} initial mass function, and adopting the main-sequence lifetime of stars from \citet{hur00} and the initial-final mass relations for stellar remnants from \citet{che90}. We consider mass loss due to stripping by the galactic tidal field according to
\begin{equation}
\frac{dM}{dt}=-\frac{M}{t_{\mathrm{tid}}}
\end{equation}
where
\begin{equation}
t_{\mathrm{tid}}(r,M)\approx 10 \left(\frac{M}{2\times 10^5\,\mathrm{M}_{\odot}}\right)^{2/3} P(r)\ \mathrm{Gyr}
\label{eqn:ttid}
\end{equation}
is the typical tidal disruption time \citep{gie08}, and
\begin{equation}
P(r)=100\left(\frac{r}{\mathrm{kpc}}\right)\left(\frac{V_{\mathrm{c}}(r)}{\mathrm{km}\ \mathrm{s}^{-1}}\right)^{-1}
\label{eqn:pergc}
\end{equation}
is the (normalized) rotational period of the cluster orbit, which parametrizes the strength of the local galactic field, and $V_{\mathrm{c}}(r)$ is the circular velocity at a distance $r$ from the galactic center.

We assume that the cluster is torn apart when the stellar density at a characteristic radius, such as the half-mass radius, falls below the mean local galactic density
\begin{equation}
\rho_{\mathrm{h}}<\rho_*(r)=\frac{V_{\mathrm{c}}^2(r)}{2\pi G r^2}\ ,
\label{eqn:dens}
\end{equation}
due to the adopted field stellar mass, as well as the growing mass of the nuclear star cluster (NSC). Following \citet{gne14}, we adopt the average density at the half-mass radius
\begin{equation}
\rho_h=
\begin{cases}
10^3\, \mathrm{M}_{\odot}\, \mathrm{pc}^{-3}& {\rm for\;} M \le 10^5\,\mathrm{M}_{\odot} \cr
10^3 \left(\frac{M}{10^5\, \mathrm{M}_{\odot}}\right)^2\, \mathrm{M}_{\odot}\, \mathrm{pc}^{-3}& {\rm for\;} 10^5\, \mathrm{M}_{\odot} <  M < 10^6\, \mathrm{M}_{\odot} \cr
10^5\, \mathrm{M}_{\odot}\, \mathrm{pc}^{-3}& {\rm for\;} M \ge 10^6\,\mathrm{M}_{\odot}
\end{cases}
\label{eqn:rhohcases}
\end{equation}
The cluster mass $M$ here is the current value before disruption, not the initial mass. As the NSC builds up in mass, its stellar density eventually begins to exceed the densities within the infalling GCs, which will be directly disrupted before reaching the center of the galaxy \citep[e.g.,][]{ant13}.

As in \citet{gne14}, we assume the clusters to orbit on a circular trajectory of radius $r$ and take this radius to be the time-averaged radius of the true, likely eccentric, cluster orbit. We consider the effect of dynamical friction on cluster orbits by evolving the orbital radius $r$ 
\begin{equation}
\frac{dr}{dt}=-\frac{r^2}{t_{\mathrm{df}}}\ ,
\label{eqn:dynf}
\end{equation}
where
\begin{equation}
t_{\mathrm{df}}(r,M)\approx 0.45 \left(\frac{M}{10^5\ \mathrm{M}_{\odot}}\right)^{-1}\left(\frac{r}{\mathrm{kpc}}\right)^2\left(\frac{V_{\mathrm{c}}(r)}{\mathrm{km}\ \mathrm{s}^{-1}}\right)\ \mathrm{Gyr}\ .
\end{equation}
We also include a correction for the non-zero eccentricities of the cluster orbits, $f_e=0.5$ \citep[for details see][]{jia08,gne14}.

\subsection{Andromeda potential model}
\label{subsec:andrompot}

We describe the Andromeda gravitational potential with a 3-component model $\Phi=\Phi_b+\Phi_{disk}+\Phi_{halo}$, where
\begin{itemize}
\item $\Phi_{b}$ is the contribution of a spherical bulge,
\begin{equation}
\Phi_{b}(r)=-\frac{GM_{b}}{r+a_b},
\end{equation}
with mass $M_{b}=1.9\times 10^{10}$ M$_{\odot}$ and core radius $a_b=1$ kpc;
\item $\Phi_{d}$ is the contribution of an axisymmetric disc,
\begin{equation}
\Phi_{d}(R,z)=-\frac{GM_{d}}{\sqrt(R^2+(b+\sqrt{c^2+z^2})^2)},
\end{equation}
with mass $M_{d}=8\times 10^{10}$ M$_{\odot}$, length scale $b=5$ kpc and scale height $c=1$ kpc;
\item $\Phi_{halo}$ is the contribution of a spherical dark matter halo
\begin{equation}
\Phi_{DM}(r)=-\frac{GM_{DM}\ln(1+r/r_s)}{r}.
\end{equation}
with $M_{DM}=2\times 10^{12}$ M$_{\odot}$ and length scale $r_s=35$ kpc.
\end{itemize}
The adopted parameters match the observed maximum circular velocity \citep{vand12,pat17}.

\section{Gamma-ray excess in Andromeda}
\label{sect:andromeda}

\begin{table*}
\caption{Model, spin-down ($\tau$), initial cluster mass fraction ($f_{\mathrm{GC},i}$), maximum GC mass ($M_{\max}$), gamma-ray luminosity-to-mass ratio ($L_{\gamma}/M_{GC}$), maximum luminosity of MSPs ($L_{\gamma,\rm max}$), slope of the MSP luminosity distribution ($\alpha$), mass of Andromeda bulge ($M_b$), total gamma-ray luminosity at $5$ kpc ($L_{\gamma}^{5}$).}
\centering
\begin{tabular}{lcccccccccc}
\hline\\[-2mm]
Model & $f_{\mathrm{GC},i}$ & $M_{\max}$ ($\msun$) & $L_{\gamma}/M_{GC}$ & $\tau$ (Gyr) & $L_{\gamma,\rm max}$ ($\mathrm{erg\, s}^{-1}$) & $\alpha$ & $M_b$ ($\msun$) & $L_{\gamma,5}$ ($\mathrm{erg\, s}^{-1}$)\\[1mm]
\hline\\[-2mm]
LON-EQ	& $0.0075$ & $10^7$ & Eq. \ref{eqn:lgmcl} & Prager+2017 & $10^{36}$ & $1$ & $1.9\times 10^{10}$ & $3.3\times 10^{37}$\\
LON-C		& $0.0075$ & $10^7$ & const & Prager+2017 & $10^{36}$ & $1$ & $1.9\times 10^{10}$ & $3.3\times 10^{37}$\\
GAU-EQ	& $0.0075$ & $10^7$ & Eq. \ref{eqn:lgmcl} & Freire+2001 & $10^{36}$ & $1$ & $1.9\times 10^{10}$ & $1.4\times 10^{37}$\\
GAU-C		& $0.0075$ & $10^7$ & const & Freire+2001 & $10^{36}$ & $1$ & $1.9\times 10^{10}$ & $1.4\times 10^{37}$\\
FGCI		& $0.0075$-$0.06$ & $10^7$ & const & Prager+2017 & $10^{36}$ & $1$ & $1.9\times 10^{10}$ & $3.3$-$21\times 10^{37}$\\
MMAX		& $0.0075$ & $0.5$-$3\times 10^7$ & const & Prager+2017 & $10^{36}$ & $1$ & $1.9\times 10^{10}$ & $2.9$-$3.6\times 10^{37}$\\
LMAX		& $0.0075$ & $10^7$ & const & Prager+2017 & $10^{35}$-$10^{36}$ & $1$ & $1.9\times 10^{10}$ & $2.9$-$3.3\times 10^{37}$\\
ALPHA		& $0.0075$ & $10^7$ & const & Prager+2017 & $10^{36}$ & $0.5$-$1.5$ & $1.9\times 10^{10}$ & $3.1$-$3.5\times 10^{37}$\\
MBUL		& $0.0075$ & $10^7$ & const & Prager+2017 & $10^{36}$ & $1$ & $0.5$-$4\times 10^{10}$ & $1.9$-$3.6\times 10^{37}$\\
\hline
\end{tabular} 
\label{tab:models} 
\end{table*}

In our model, everything has been fixed apart from the initial amount of galactic mass locked in GCs. The initial cluster mass fraction $f_{\mathrm{GC},i}$ is generally of the order of a few percent, but its exact value is difficult to estimate. In the case of the Milky Way, it can be fixed by assuming that a certain fraction of the NSC was accreted by inspiralling GCs \citep{fao18}. To overcome this problem, we make use of a strong correlation between the present-day mass of the GC population and the host halo that emerges both from observations and models \citep*[e.g.,][]{harris2017,cho18}
\begin{equation}
M_{GC}=3.4\times 10^{-5}\ M_{DM}\ .
\label{eqn:mgcmh}
\end{equation}
with an intrinsic scatter of $\sigma=0.2$ dex. Thus we set $f_{\mathrm{GC},i}=0.0075$, which gives a final present-day mass of the GC system in Andromeda which agrees within $1\sigma$ with equation~(\ref{eqn:mgcmh}).

We evolve the GC population according to the model in \S\ref{sect:gcev} and compute the mass deposited by each cluster as a function of time $t$ and radius $r$ from Andromeda's galactic centre. Then we calculate the total amount of gamma-ray luminosity expected from all MSPs left in the cluster debris, by using the mean relation between the gamma-ray emission from GCs and their masses \citep{fao18}
\begin{equation}
\log\frac{L_\gamma}{M_\mathrm{GC}}=32.66\pm 0.06 -(0.63\pm 0.11) \log{M_\mathrm{GC}} \,,
\label{eqn:lgmcl}
\end{equation}
where $L_\gamma$ is the gamma-ray emission of a GC in erg~s$^{-1}$, and $M_\mathrm{GC}$ is its mass in units of $M_\odot$. Alternatively, we also consider models with 
\begin{equation}
\frac{L_\gamma}{M_\mathrm{GC}} = \mathrm{const}\, = 4.57\times 10^{29}~\mathrm{erg~s}^{-1} M_\odot^{-1}\ ,
\label{eqn:lgmclcon}
\end{equation}
to test the dependence of our results on the adopted $L_{\gamma}-M_{GC}$ relation. We then generate individual MSPs by sampling from a power-law distribution
\begin{equation}
  \frac{dN}{dL_{\gamma}}\propto L_{\gamma}^{-\alpha}\ ,
  \label{eqn:lgdist}
\end{equation}
with $\alpha=1$, between $L_{\gamma,\rm min}=10^{31}\,\mathrm{erg\, s}^{-1}$ and $L_{\gamma,\rm max}=10^{36}\,\mathrm{erg\, s}^{-1}$ \citep{ajel16}. We sample from the above distribution until the total luminosity from the deposited MSPs equals $L_{\gamma,\rm tot}^{\rm dep}(t)$. This gives us the number of MSPs, $N_{\rm MSP}(t)$.

\begin{figure} 
\centering
\includegraphics[scale=0.55]{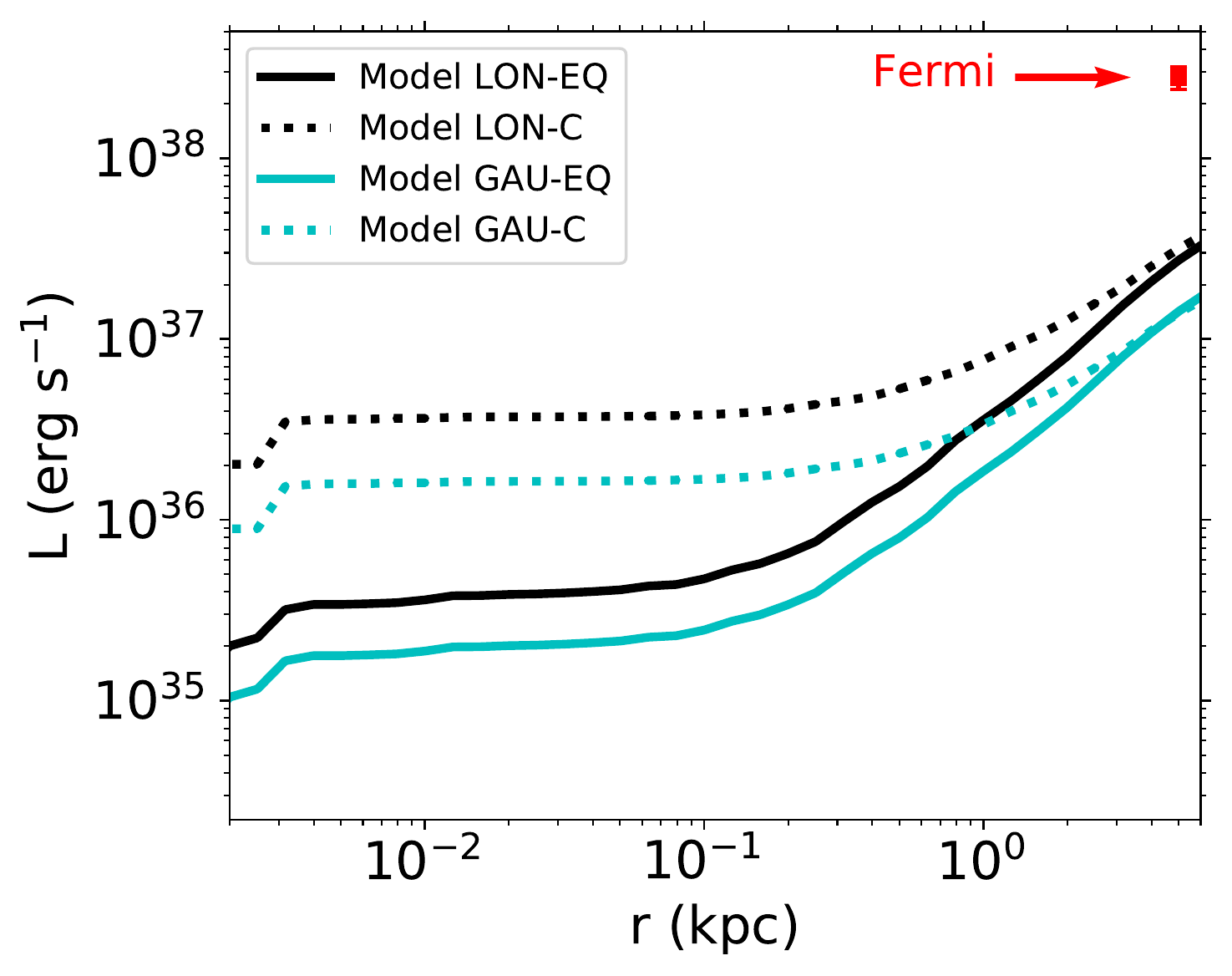}
\caption{Predicted MSP integrated gamma-ray luminosity within a distance $r$ from Andromeda's centre. The red data-point represents the luminosity as measured by \textit{Fermi}.}
\label{fig:excess}
\end{figure}

From the moment that MSPs are deposited in the galactic centre, we evolve in time the gamma-ray luminosity of a given pulsar as
\begin{equation}
L_{\gamma}(t)=\frac{L_{\gamma,0}}{[1+(t/\tau)^{1/2}]^2} ,
\label{eqn:lgamma}
\end{equation}
where $L_{\gamma,0}$ is the initial luminosity and $\tau$ is the characteristic spin-down timescale for a MSP to lose its rotational kinetic energy due dipole magnetic breaking
\begin{equation}
\tau=\frac{E}{\dot{E}} = \frac{P}{2\dot{P}},
\label{eqn:tau}
\end{equation}
where $P$ and $\dot{P}$ are the MSP rotational period and its derivative, respectively. As discussed in \citet{fao18}, we adopt two models for the MSP spin-down. The first model (Model LON) uses observations of the MSP population in 47 Tuc, and $\tau$ is given by \citep{prager17}
\begin{equation}
\tau=\frac{c}{1.59\times 10^{-9} \mathrm{m\ s^{-2}}} \left(\frac{2\times 10^8\ \mathrm{G}}{B}\right)^2 \left(\frac{P}{2\ \mathrm{ms}}\right)^2\ ,
\label{eqn:pdot}
\end{equation}
where $c$ is the speed of light and $B$ is the magnetic field. In this model, the $\tau$ distribution has a mean around $1$~Gyr, but also a non-negligible tail at larger $\tau$'s. In the second model (Model GAU), we adopt a Gaussian distribution with mean of $3\,$Gyr, consistent with \citet{freir01}, who found a characteristic age of $\approx 3\,$Gyr for MSPs in NGC 104. We note that recently \citet{ole16} have claimed that $L_\gamma\propto (1+(t/\tau)^{1/2})^{-1}$ is more consistent with the data, which would give a less important spin-down of MSP luminosities. The shape of the $\tau$ distribution and its relation to $L_\gamma$ turn out to be the two most important ingredients controlling the final contribution of the excess, but both of them are still quite uncertain \citep{fao18}. Table \ref{tab:models} summarizes the models considered in the present work.

In Figure \ref{fig:excess}, we illustrate the predicted MSP integrated gamma-ray luminosity at $2$~GeV within a distance $r$ from Andromeda's centre. We found that the total gamma-ray luminosity is $\sim 1.6\times 10^{37}$ erg s$^{-1}$ and $\sim 3.5\times 10^{37}$ erg s$^{-1}$ for Model GAU-EQ and Model LON-EQ, respectively. As already noted in \citet{fao18}, we found that the models that use the \citet{prager17} prescription for the spin-down rate predict a flux about two times larger than the models with the \citet{freir01} $\tau$ distribution. We also ran models with a constant value of gamma-ray luminosity to mass ratio to compute the total amount of gamma-ray flux in cluster debris. In these latter models, the overall flux at $r\lesssim 1$ kpc is increased by about an order of magnitude, but the total luminosity is comparable.

\subsection{Model uncertainties}
\label{sect:uncertainties}

\begin{figure} 
\centering
\includegraphics[scale=0.55]{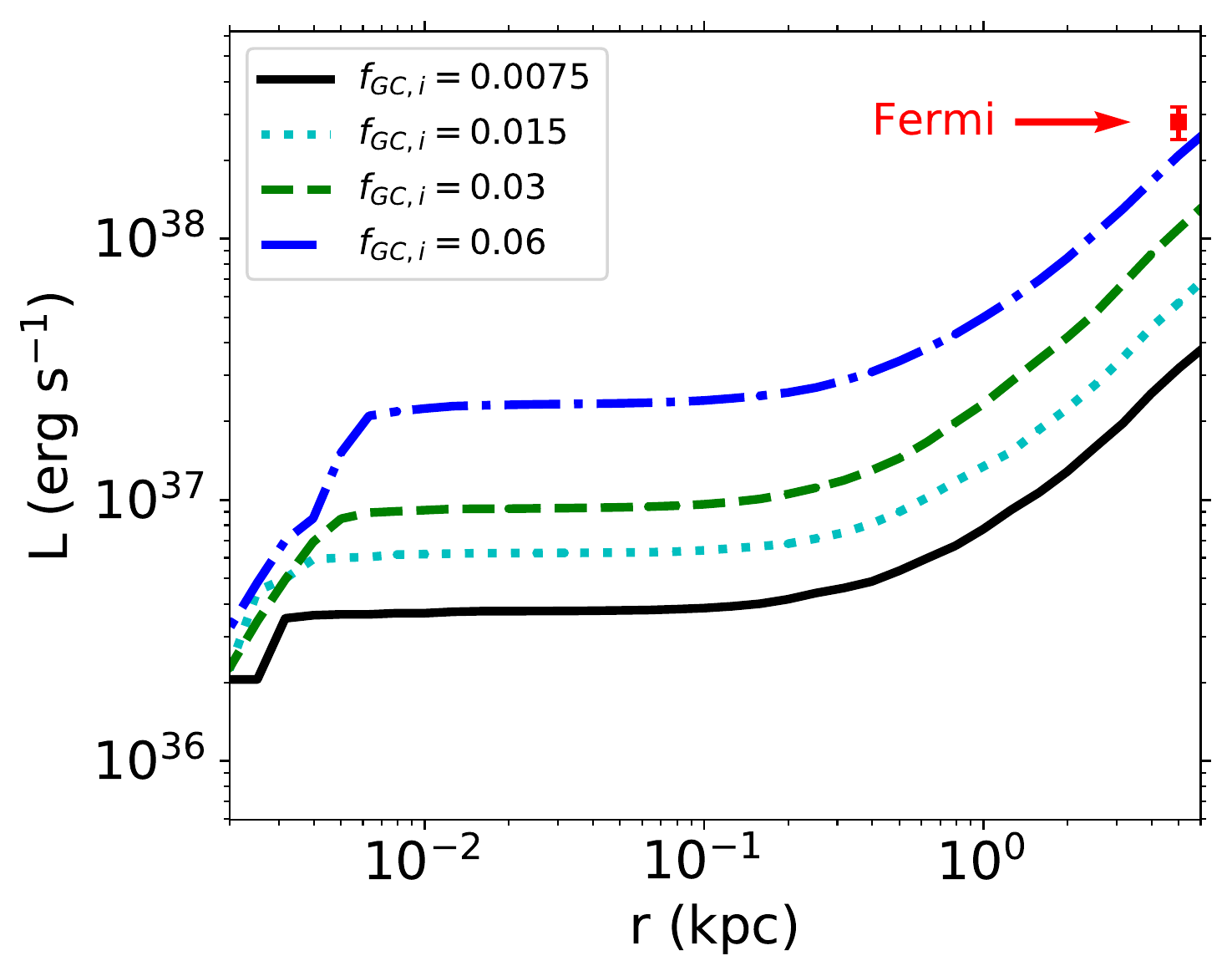}
\caption{Predicted MSP integrated gamma-ray luminosity within a distance $r$ from Andromeda's centre as function of the initial amount of mass in GCs (Model FGCI). The red data-point represents the luminosity as measured by \textit{Fermi}.}
\label{fig:excess_fgci}
\end{figure}

In our fiducial model LON-C, the MSP scenario does not reproduce the observed gamma-ray flux measured at the centre of Andromeda. Yet, there are some important uncertainties in our approach, apart from the MSP spin-down model \citep[see also discussion in][]{fao18}. 

An important factor that can affect the results is the initial cluster mass fraction. In the models described above, we fixed $f_{\mathrm{GC},i}$ by requiring that the present-day mass of the GC system agrees with the cosmological scaling relation (Eq.~\ref{eqn:mgcmh}). To check the effect of this parameter, we ran Model FGCI where we consider $0.0075\le f_{\mathrm{GC},i}\le 0.06$. The other parameters are set as in the fiducial Model LON-C. Figure \ref{fig:excess_fgci} shows the resulting MSP integrated gamma-ray luminosity within a distance $r$ from Andromeda's centre as a function of the initial amount of mass in GCs. Our MSP model can reproduce the observed \textit{Fermi} excess only for $f_{\mathrm{GC},i}\approx 0.06$; this is $\sim 8$ larger than the $f_{\mathrm{GC},i}$ inferred from Eq.~\ref{eqn:mgcmh} and predicts a larger number of clusters in Andromeda than expected \citep{barm01}.

In our main model, we fixed $M_{\min}=10^4\msun$ and $M_{\max}=10^7\msun$. Our results are essentially independent on $M_{\min}$, because low-mass clusters dissolve quickly \citep{gne14}. We explore the effect of varying $M_{\max}$ on our results (Model MMAX; see Tab.~\ref{tab:models}). We find that a larger $M_{\max}$ implies a larger gamma-ray excess, but only by a small factor. More specifically, we find that the integrated MSP gamma-ray luminosity is $\sim 2.9\times 10^{37}$ erg~s$^{-1}$ and $\sim 3.6\times 10^{37}$ erg~s$^{-1}$ for $M_{\max}=5\times 10^6\msun$ and $M_{\max}=3\times 10^7\msun$, respectively. These results demonstrate that the integrated gamma-ray luminosity predicted by our models is only marginally affected by the choice of $M_{\max}$.

Another source of uncertainty is the parametrization of the typical timescales for cluster evolution and $\rho_h$. In case of Eq.~\ref{eqn:rhohcases}, we take the lower limit from the typical observed density of low-mass Galactic GCs. More massive clusters are expected to be in the expansion phase to fill their Roche lobes, during which $\rho_h \propto M^2$ \citep{gieles_etal11}. The upper limit for the most massive clusters ($\rho_{\mathrm{h}} = 10^5\ \mathrm{M}_{\odot}$ pc$^{-3}$) corresponds roughly to the highest observed half-mass density of Galactic GCs. For what concerns Eq.~\ref{eqn:pergc}, we have revised the normalization of $P(r)$, hence of Eq.~\ref{eqn:ttid}, relative to our first paper \citep{fao18} by a factor of $\sim 2.5$, to account for the longer disruption time in detailed N-body simulations of \citet{lamers2010}. However, by comparing our current models to the ones presented in \citet{fao18}, we found no significant effect on the results.

Finally, we do not consider eccentric GC orbits in our model, rather we include the effect of the deviation of the cluster's orbit from circular by taking into account a correction factor $f_e=0.5$ in Eq.~\ref{eqn:dynf}, consistent with the results of simulations by \citet{jia08}. We note that eccentric orbits may have shorter dynamical friction timescales, increase the mass-loss rate, and shorten the GC relaxation time. Thus, some of the clusters may get disrupted earlier than clusters on circular orbits. Unfortunately, the primordial distribution of cluster eccentricities is not known, which makes it difficult to quantify its effect. However, we note that the simple prescriptions for dynamical friction as implemented here reproduce well the spatial and mass distribution of GCs in our Galaxy \citep{gne14}.

Also the parameters defining the MSP luminosity distribution can play a role. In our fiducial model (Model LON-C), we set the slope of the distribution $\alpha=1$, and $L_{\gamma,\rm min}=10^{31}\,\mathrm{erg\, s}^{-1}$ and $L_{\gamma,\rm max}=10^{36}\,\mathrm{erg\, s}^{-1}$. While $L_{\gamma,\rm min}$ does not play a significant role \citep{fao18}, we investigate the impact of $L_{\gamma,\rm max}$ and $\alpha$ on our results (see Tab.~\ref{tab:models}). We repeated our calculations with $L_{\gamma,\rm max}$ reduced to $10^{35}\ \mathrm{erg\, s}^{-1}$ (Model LMAX), and found that the total gamma-ray excess is reduced by $\sim 10\%$ with respect to our fiducial model. Also, we considered $\alpha$ in the range $0.5$-$1.5$ (Model ALPHA). Smaller $\alpha's$ (shallower distributions) imply a larger gamma-ray emission, while larger $\alpha's$ (steeper distributions) lead to a smaller total excess. In any case, the Andromeda excess results $\sim 5\%$ bigger and $\sim 5\%$ smaller for $\alpha=0.5$ and $\alpha=1.5$, respectively, than our fiducial model.

We also check the effect of varying the parameters defining the Andromeda bulge mass $M_b$, which may be an important parameter for the GC survivability in the galaxy innermost regions. We find that the integrated gamma-ray excess is $\sim 1.9\times 10^{37}$ erg s$^{-1}$ and $\sim 3.6\times 10^{37}$ erg s$^{-1}$ when $M_b=0.5\times 10^{10}$ and $M_b=4\times 10^{10}$, respectively, and still smaller than the observed one. Finally, we also ran models adopting different values of the Andromeda bulge scale radius $a_b$, but we did not find any significant difference in the total gamma-ray flux compared to our fiducial Model LON-C.

Table~\ref{tab:models} reports all the models we investigated and their total gamma-ray luminosity within $5$ kpc ($L_{\gamma,5}$).

\section{Conclusions}
\label{sect:conclusions}

The \textit{Fermi} Telescope has revealed a gamma-ray excess around our Galactic Center (out to $\sim 3\,$kpc) of the order of $\sim 10^{37}$ erg s$^{-1}$, which has been interpreted either as dark matter annihilation emission or as emission of thousands of MSPs. \textit{Fermi} also showed evidence of a diffuse gamma-ray emission ($\sim 2.8 \times 10^{38}$ erg s$^{-1}$) also in the centre (up to $\sim 5$ kpc) of the Andromeda galaxy. As in the case of the Galactic Centre, there have been suggestions for both a MSP and for a dark matter-annihilation emission. 

In this letter, we have revisited the MSP scenario in the Andromeda galaxy, by modeling the formation and disruption of GCs, that can deliver thousands of MSPs in the bulge. We have modeled the MSP gamma-ray emission by taking into account also the spin-down due to magnetic-dipole braking, and found that the total gamma-ray luminosity is $\sim 1.6$-$3.5\times 10^{37}$, i.e., nearly one order of magnitude smaller than the observed excess. Our MSP model can reproduce the \textit{Fermi} excess only by assuming a number of primordial clusters that is  $\sim 8$ times larger than that inferred from the galactic scaling relations. 

Recently, \citet{eckn17} proposed that the emission from an unresolved population of MSPs formed \textit{in situ} can account for $\sim 7 \times 10^{37}$ erg s$^{-1}$ of the excess. While both our model and the \citet{eckn17} model cannot account for all the observed excess, they can explain nearly half of it when taken together. We also note that M31 likely had a burst of star formation $\sim 1-2$ Gyr ago, which could boost both the abundance of close binaries and massive star clusters up to a factor of $\sim 2$ \citep{dong2018}. A combination of all these factors could provide the astrophysical origin of the gamma-ray emission in the Andromeda galaxy.

Finally, we note that some of the neutron stars delivered by the GCs may mass-segregate to some extent and also be successfully exchanged in few-body interactions in binaries that later could lead to the formation of MSPs, which could enhance our predicted rate \citep{leiant2016}. This would give a maximum contribution roughly comparable to the \textit{in situ} formation scenario, which can account only for $\sim 1/4$ of the excess \citep{eckn17}, being the mass in GCs of the order of the mass of the nuclear star cluster. However, the details of binary modeling would be the same for Andromeda and our Galaxy, while the observed gamma-ray fluxes are very different. Hence, MSPs delivered by GCs cannot explain both the Milky Way and Andromeda fluxes, and therefore other sources of gamma-rays in M31 center are required.

\section{Acknowledgements}

GF is supported by the Foreign Postdoctoral Fellowship Program of the Israel Academy of Sciences and Humanities. GF also acknowledges support from an Arskin postdoctoral fellowship at the Hebrew University of Jerusalem. FA acknowledges support from an E. Rutherford fellowship (ST/P00492X/1) from the Science and Technology Facilities Council. OG was supported in part by the NSF through grant 1412144.

\bibliographystyle{yahapj}

\end{document}